\documentclass[aps,pre,12pt,showkeys]{revtex4-1}
\usepackage{graphicx}
\usepackage{epstopdf}
\usepackage{amsmath}
\usepackage{amssymb}
\usepackage{default}
\usepackage{natbib}
\newcommand{\be}{\begin{equation}}
\newcommand{\bea}{\begin{eqnarray}}
\newcommand{\bc}{\begin{center}}            
\newcommand{\ee}{\end{equation}}
\newcommand{\eea}{\end{eqnarray}}
\newcommand{\ec}{\end{center}}

\newcommand{\baa}{\begin{eqnarray*}}
\newcommand{\eaa}{\end{eqnarray*}}
\begin{document}
\title{Low-dissipation Carnot-like heat engines at maximum efficient power}
\author{Varinder Singh}
\email{varindersingh@iisermohali.ac.in}
\author{Ramandeep S. Johal}
\email{rsjohal@iisermohali.ac.in}
\affiliation{ Department of Physical Sciences, \\ 
Indian Institute of Science Education and Research Mohali,
Sector 81, S.A.S. Nagar, Manauli PO 140306, Punjab, India}
\begin{abstract}
We study the optimal performance of Carnot-like heat engines working in low 
dissipation regime using the product of the efficiency and the power output,
also known
as the efficient power, as our objective function. 
Efficient power function 
represents the best trade-off between power and efficiency of a heat engine. We 
find lower and upper bounds on the efficiency in case of extreme 
asymmetric dissipation when the ratio of dissipation coefficients at the 
cold and the hot contacts approaches, respectively, zero or infinity. 
In addition, we obtain the form of efficiency for the case of symmetric 
dissipation.
We also discuss the universal features of efficiency at maximum efficient
power
and derive the bounds on the efficiency using global linear-irreversible 
framework
introduced recently by one of the authors. 
\end{abstract} 
\maketitle
\section{Introduction}
Carnot efficiency,  $\eta_{\rm C}=1-T_c/T_h$, sets a theoretical upper 
bound on the efficiency of 
all heat engines working between two heat baths at temperatures $T_c$ 
and $T_c$ ($T_c<T_h$). 
The Carnot efficiency is attainable only in the reversible limit, whereby 
the processes occur so slowly that the  
resulting output power is zero. But, real heat engines operate at finite 
rates and hence produce finite power per cycle. So it is more useful to 
optimize the power output of the heat engines. The derivation of 
Curzon-Ahlborn (CA) 
efficiency $\eta_{CA}=1-\sqrt{1-\eta_{\rm C}}$ of an endoreversible 
engine \cite{Curzon1975}, 
operating at maximum power (MP), marked the beginning of finite-time 
thermodynamics (FTT) 
\cite{devosbook,Berrybook,WuChenbook}. 
In endoreversible models \cite{Curzon1975,devos1985,Rubin}, 
the work extracting part of the cycle is assumed
to be internally reversible and there are no heat leaks between the heat baths. 
The irreversibility arises solely due to the finite rate of heat transfer 
between the 
working medium and the external heat baths. However, CA efficiency is not a 
universal result, and it is neither an upper nor a lower bound 
\cite{Broeck2005}. 
In the linear response regime, efficiency at maximum power (EMP) comes out to be
$\eta_{\rm C}/2$ for the the tight coupling condition \cite{Broeck2005}. At the 
level of nonlinear response,
Esposito \textit{et al.} proved that second order term $\eta_{\rm C}^2/8$ is 
also universal 
if we have a left-right symmetry in addition to tight coupling condition 
\cite{Lindenberg2009}.

Recently, using the assumption of low dissipation (LD), Esposito \textit{et al} 
\cite{Esposito2010}, 
derived upper and lower bounds for the EMP of Carnot-like heat engines. 
In addition, for the symmetric dissipation, they were able to 
reproduce the CA result. The LD models 
\cite{Esposito2010,Broeck2013,Schmiedl2008,deTomas2012,WangLi2012,deTomas2013,
Guo2013,
Holubec2015,Holubec2016,Hernandez2015,WangHe2012,WangHeWu2012,GuoWang2013,
HuWu2013} 
have some advantages over the endoreversible models. It does not make use of 
any 
specific heat transfer law and also valid beyond the linear-response regime. A 
good comparison of
LD models and endoreversible models is given in the Refs. 
\cite{Johal2017,Ayala2016,Ayala2017,Tang2018}.
Further, LD models were used to investigate the optimal performance of 
Carnot-like
refrigerators \cite{deTomas2012,WangLi2012,HuWu2013}, quantum heat engines 
\cite{WangHeWu2012,GuoWang2013} 
and for the optimization of target functions other than power output 
\cite{deTomas2013,Holubec2015,Holubec2016}. Guo et. al. investigated the 
the optimal performance of LD heat engines for different types of heat cycles 
other than Carnot cycle \cite{Guo2013}.

But, heat engines operating at MP are not the most efficient ones 
and, hence, are not much economical. It has been already pointed out that 
actual thermal plants and heat engines should not operate at MP, 
but in a regime with slightly smaller power and appreciable larger efficiency 
\cite{Chen2001,devosbook}. 
The optimization of Omega criterion or ecological criterion 
\cite{ABrown1991,Hernandez2001,VarinderJohal} and efficient power 
criterion \cite{Stucki,YanChen1996,Yilmaz} falls in such a regime. They pay 
equal attention to both 
power output and efficiency \cite{Arias2009}. In this work, we investigate the 
optimization of efficient 
power criterion for a Carnot-like engine working in LD regime. 

Efficient power criterion $P_{\eta}=\eta P$ represents the best compromise 
between the efficiency and power output of a heat engine. In was introduced 
by Stucki \cite{Stucki} in the context of linear-irreversible thermodynamics 
(LIT) while 
studying the mitochondrial energetic processes. Later the idea was extended 
to the regime of FTT by Yan and Chen \cite{YanChen1996} and given the so-called 
name 
efficient power by Yilmaz \cite{Yilmaz}. It is also shown that the efficient 
power criterion is also well suited to study the optimization of steady and 
non-steady 
electric energy converters \cite{Valencia2016}, thermionic generator 
\cite{ChenDing} and biological systems
\cite{Stucki,ABrown2008,Chimal}. For some naturally designed biological 
systems, 
maximum efficient power (MEP) conditions may lead to more efficient
engines than those at maximum Omega function (MOF) or ecological function 
\cite{Chimal}. 

In this paper, we analyse the optimal performance of general class of LD 
Carnot-like heat engines
using efficient power function as the objective function. In Sec.II, we discuss 
model of LD heat 
engine undergoing Carnot cycle. In Sec. III, we find the general expression for 
EMEP and 
obtain lower and upper bound on the efficiency. We also discuss universal 
features of
EMEP in this section. Sec. IV is devoted to the comparison of rates of 
dissipation at hot 
and cold contacts for three different objective functions. In Sec. V, using a 
different optimization scheme,
we obtain the same bounds on the EMEP as obtained for LD heat engines.
We conclude in Sec. VI by highlighting the key results.
\section{Model of low-dissipation Carnot engine}
As in the case of usual Carnot cycle, heat cycle in our case consists of two 
adiabatic and two isothermal steps. Adiabatic steps are assumed to be 
instantaneous 
and there is no entropy production along these branches. Let $t_h$ and $t_c$ be 
the 
time durations of the isothermal branches during which the system remains in 
contact 
with the hot and cold reservoirs respectively. During the heat exchange process 
with the hot (cold) bath,
the change in entropy of the system can be split into two parts as follows
\begin{equation}
\Delta S_j = \Delta S_j^{r} + \Delta S_j^{ir},\qquad j=h, c
\end{equation}
where $\Delta S_j^{r}$ is change in entropy of the system due to reversible heat 
transfer 
and $\Delta S_j^{ir}$ accounts for irreversible entropy production during the 
process. The 
first term is $Q_h/T_h$ for the heat absorbed from the hot reservoir at 
temperature 
$T_h$ and $Q_c/T_c$ for the heat transferred to cold reservoir at temperature 
$T_c$. 
In low dissipation limit, it is assumed that irreversible entropy production 
$\Delta S_j^{ir}$ 
during the heat transfer step is inversely proportional to the  time duration 
for which the system 
remains in contact with the bath. Hence entropy production along the isothermal 
branch is given by 
$\Delta S_j^{ir} = \Sigma_j/t_j$, ($j=h,c$). Here $\Sigma_h$ and $\Sigma_c$ are 
dissipation 
coefficients, containing the information about the irreversibilities induced in 
the model as we 
deviate away from the reversible limit. It is self evident that the cycle 
approaches reversible 
limit as $t_h\rightarrow\infty$ and $t_c\rightarrow\infty$. Thus, at hold and 
cold contacts, 
we have respectively
\begin{eqnarray}
\Delta S_h &=& \frac{Q_h}{T_h} + \frac{\Sigma_h}{t_h},
\\
\Delta S_c &=& -\frac{Q_c}{T_c} + \frac{\Sigma_c}{t_c},
\end{eqnarray}
where $Q_h,Q_c>0$. Since after undergoing the full cycle, the system returns to 
its initial state, the total entropy change in the whole cycle is zero: $\Delta 
S_h+\Delta S_c=0$. 
Therefore we have $\Delta S_h=-\Delta S_c=\Delta S>0$. Then the amount of heat 
exchanged with each 
reservoir can be written as
\begin{eqnarray}
Q_h &=& T_h\left(\Delta S-\frac{\Sigma_h}{t_h}\right)\equiv T_h(\Delta 
S-x_h\Sigma_h),\label{heat1}
\\
Q_c &=& T_c\left(\Delta S+\frac{\Sigma_c}{t_c}\right)\equiv T_c(\Delta 
S+x_c\Sigma_c),\label{heat2}
\end{eqnarray}
where we have used $x_h\equiv 1/t_h$ and $x_c\equiv 1/t_h$ for our convenience.
The work extracted in a cycle with time period $t=t_c+t_h$ is $W=Q_h-Q_c$. So 
the efficiency $\eta$ 
and average output power $P$ per cycle is defined as
\begin{equation}
\eta = \frac{W}{Q_h}=1-\frac{Q_h}{Q_c}=1-\frac{T_c(\Delta 
S+x_c\Sigma_c)}{T_h(\Delta S-x_h\Sigma_h}, \label{etaa}
\end{equation}
\begin{equation}
P = \frac{Q_h-Q_c}{t_h+t_c} \equiv \frac{(Q_h-Q_c)x_h x_c}{x_c+x_h} 
\label{powerr}
\end{equation}
\section{Efficient power in low dissipation regime}
To study the optimal performance of a low dissipation Carnot engine, we will use 
efficient power  
$P_{\eta}=\eta P$ as the target function. Here, the efficient power represents 
the best compromise 
between the efficiency and average power of the engine. Using Eqs. (\ref{etaa}) 
and (\ref{powerr}) in 
the expression for $P_{\eta}$, we have
\begin{equation}
P_{\eta} = \eta P = \frac{(Q_h - Q_c)^2}{Q_h } \frac{  x_c x_h}{x_c + x_h}.
\end{equation}
Setting the partial derivatives of $P_{\eta}$ with respect to $x_c$ and $x_h$
equal to zero, we have respectively the following two equations:
\begin{equation}
\frac{(Q_h - Q_c)^2}{Q_h } x_h = 2 T_c\Sigma_c (x_c + x_h)x_c  
\left[ 1 - \frac{ T_c(\Delta S+x_c\Sigma_c)}{T_h(\Delta S-x_h\Sigma_h)}\right],
\label{opti1}
\end{equation} 
and
\begin{equation}
\frac{(Q_h - Q_c)^2}{Q_h }x_c 
=
 T_h\Sigma_h (x_c + x_h)x_h 
 \left[ 1 -\frac{T_c^2}{T_h^2}\frac{ (\Delta S+x_c\Sigma_c)^2}{(\Delta 
S-x_h\Sigma_h)^2}\right].
\label{opti2}
\end{equation}
Using Eqs. (\ref{heat1}) and (\ref{heat2}) in Eqs. (\ref{opti1}) and 
(\ref{opti2}), we solve for $x_h$ and 
get the following expression (see Appendix A) for $x_h$
\begin{equation}
x_h = -\frac{\Delta S N}{8\Sigma_h B}-\frac{1}{2}\sqrt{K}-\frac{1}{2} 
\sqrt{K'-\frac{1}{4\sqrt{K}}      
\left[ \frac{\Delta S^3}{\Sigma_h^3} 
\left(  \frac{12\eta_{\rm C} \Sigma_h}{B} -\frac{N^3}{8 B^3} + \frac{M N}{2 B^2} 
  \right)       
\right]} 
\label{xh1}
\end{equation}
where we used the following notation 
\begin{eqnarray}
K &=& \frac{ \Delta S^2}{\Sigma_h^2} \left[   \frac{N^2}{16  B^2} - \frac{M}{6 
B} 
+ 
\frac{(A+\sqrt{A^2-4 A'^3})^{1/3}}{12\times 2^{1/3}\Sigma_h B}  
+
\frac{\Sigma_h T}{6\times 2^{2/3} B (A+\sqrt{A^2-4 A'^3})^{1/3}}  \right] 
\nonumber,
\\
K' &=& \frac{\Delta S^2}{\Sigma_h^2} \left[   \frac{N^2}{8 B^2} - \frac{M}{3 B}
- 
\frac{(A+\sqrt{A^2-4 A'^3})^{1/3}}{12\times 2^{1/3}\Sigma_h B}
-
\frac{\Sigma_h T}{6\times 2^{2/3} B (A+\sqrt{A^2-4 A'^3})^{1/3}} 
\right]\nonumber ,
\\
N &=& \Sigma_h\left[(1-\eta_{\rm C})(6-\eta_{\rm C}) \gamma - 6 \right], 
\nonumber
\\
B &=& \Sigma_h \left[-(1-\eta_{\rm C})\gamma + 1 \right], \nonumber 
\\
M &=& \Sigma_h\left[-3(1-\eta_{\rm C})(3-\eta_{\rm C})\gamma + (4\eta_{\rm C}+9) 
\right],\nonumber
\\
T &=& \Sigma_h^2[9(1-\eta_{\rm C})^2(3-\eta_{\rm C})^2\gamma^2 
-
6(1-\eta_{\rm C})(3-2\eta_{\rm C})(9-5\eta_{\rm C})\gamma
+ (9-8\eta_{\rm C})^2 ],\nonumber
\\
A &=& 2 M^3 \Sigma_h^3 + 108 \eta_{\rm C} \Sigma_h^4 M N + 108 \eta_{\rm C}^2 
\Sigma_h^4 N^2
+ 
3888\eta_{\rm C}^2 \Sigma_h^5 B -288 \eta_{\rm C}^2  \Sigma_h^4 M B \nonumber ,
\\
A' &=& M^2 \Sigma_h^2 + 36\eta_{\rm C}  \Sigma_h^3 N + 48 \eta_{\rm C}^2  
\Sigma_h^3 B.\label{constants}
\end{eqnarray}
In the above equations, we have introduced the parameter 
$\gamma=\Sigma_c/\Sigma_h$.
Now we seek the form of efficiency at maximum efficienct power (EMEP)
$\eta^*=W/Q_h$, which is found to be (see Appendix B)
\begin{equation}
\eta^* = \frac{2\eta_{\rm C}}{3-2x_h\Sigma_h /\Delta S}.
\label{eff2}
\end{equation}
Using Eqs. (\ref{xh1}) and (\ref{constants}) in Eq. (\ref{eff2}), we can obtain 
a closed-form expression for EMEP.
The resulting form is too lengthy to be reproduced here. 
However, a couple of points about this expression
need to be noted. 
Firstly, it depends only upon Carnot efficiency $\eta_{\rm C}$ and 
parameter $\gamma$. For some limiting cases, it reduces to well known 
forms for the efficiency obtained in literature. 
In the extreme asymmetric limit $\gamma\rightarrow 0$, the
EMEP converges to the upper bound  $\eta_+=(3-\sqrt{9-8\eta_{\rm C}})/2$,
while for $\gamma\rightarrow\infty$, it reduces 
to the lower bound $\eta_-=2\eta_{\rm C}/3$. Thus 
\begin{equation}
\eta_-\equiv \frac{2}{3}\eta_{\rm C} \leq \eta^* \leq 
\frac{1}{2}(3-\sqrt{9-8\eta_{\rm C}})\equiv \eta_+ . \label{bounds}
\end{equation}
These upper and lower bounds on the efficiency were previously obtained by 
Holubec and Ryabov \cite{Holubec2015}
for the case of overdamped brownian particle undergoing a Carnot-like cycle 
using the framework of 
stochastic thermodynamics \cite{Schmiedl2008}.

We pay special attention to the case of symmetric dissipation in which 
$\Sigma_c=\Sigma_h$,
or $\gamma=1$. Under this condition, Eq. (\ref{eff2}) reduces to
\begin{equation}
\eta_{\rm sym} = 1 - \frac{1}{4} (1-\eta_{\rm C})\left(1+\sqrt{1 + 
\frac{8}{1-\eta_{\rm C}}}\right).\label{etasym}
\end{equation}
The same result was obtained in Refs. \cite{YanChen1996,Yilmaz} 
for the endoreversible model of Carnot heat engine
operating at MEP, under the tight-coupling condition. We expand Eq. 
(\ref{etasym}) in 
Taylor's series near equilibrium to reveal universal features of the EMEP. 
\begin{equation}
\eta_{sym} = \frac{2}{3} \eta_{\rm C} + \frac{2}{27} \eta_{\rm C}^2 + 
O(\eta_{\rm C}^3).
\end{equation}
\begin{figure}[ht]
 \begin{center}
\includegraphics[width=8.6cm]{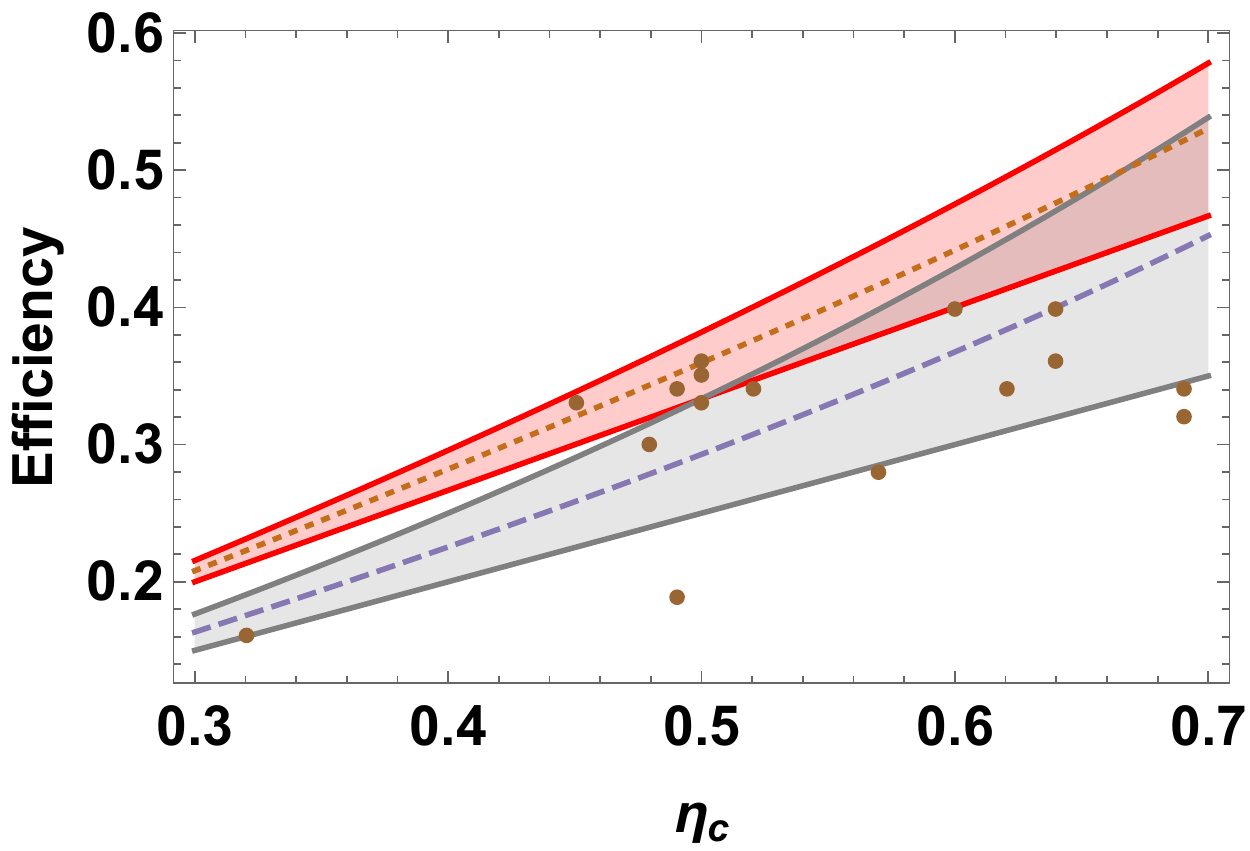}
 \end{center}
\caption{(Color online) Comparison of the bounds on efficiency with observed 
data. 
Red curves show the bounds for the EMEP. Gray lines represents the same for the 
EMP \cite{Esposito2010}.
Brown circles represent the observed efficiencies of various power plants as 
analyzed in Refs. 
\cite{Esposito2010,Holubec2015,Johal17}. Dashed and dotted lines stand for 
$\eta_{\rm CA}$ 
and $\eta_{\rm sym}$ respectively.}
\end{figure}
The first two terms in the above equation were also derived for the EMEP of 
a nonlinear irreversible heat engine \cite{Zhang2017} working in strong coupling 
limit 
under the symmetric condition by using master equation model 
\cite{Lindenberg2009,Zhang2016}. 
In Ref. \cite{Zhang2017}, it is also shown that EMEP is given by $2\eta_{\rm 
C}/3$ in 
linear response regime. Hence, we confirm that universal features of 
efficiency \cite{Lindenberg2009,Zhang2016,VarinderJohal2018} are not exclusive 
to the conditions of MP and MOF but also extend to the engines operating in MEP 
regime. 
\section{Rates of dissipation}
Now we compare the average rates of dissipation for LD heat engines under 
optimal
working conditions for power output, efficient power function and Omega 
($\Omega$) function. 
In general, the average rates of dissipation for the LD model,
at hot and cold contacts are given by \cite{Johal2017}:
\begin{equation}
\mathcal{D}_{h}^{(f)} = \frac{T_h\Sigma_h}{t_h^2} \equiv T_h\Sigma_h x_h^2,
\end{equation}
 \begin{equation}
\mathcal{D}_{c}^{(f)} = \frac{T_c\Sigma_c}{t_c^2} \equiv T_c\Sigma_c x_c^2,
\end{equation}
where $f (\equiv P,\; P_\eta,\; \Omega)$ is the function being optimized. 
In case of LD engines operating at maximum power, the relation 
between $x_h$ and $x_c$ is given by \cite{Esposito2010}
\begin{equation}
\frac{x_h}{x_c} = \sqrt{\frac{T_c\Sigma_c}{T_h\Sigma_h}},
\end{equation}
from which it follows that the average rates of dissipation at two thermal 
contacts are equal:
\begin{equation}
 \mathcal{D}_{c}^{(P)}=\mathcal{D}_h^{(P)}.  \label{diss1}
\end{equation}
Similarly for the case of maximum $\Omega$ function, we have \cite{deTomas2013}
\begin{equation}
\frac{x_c}{x_h} = \sqrt{\frac{\Sigma_h(2-\eta_{\rm C})}{2\Sigma_c(1-\eta_{\rm 
C})}}. 
\end{equation}
So, we obtain 
\begin{equation}
\mathcal{D}_c^{(\Omega)} = \mathcal{D}_h^{(\Omega)} \left(1-\frac{\eta_{\rm 
C}}{2}\right).\label{diss2}
\end{equation}
Since the factor $(1-\eta_{\rm C}/2)$ is always smaller than 1, the rate of 
dissipation 
is higher at the hot contact. Now we find the relation between rates of 
dissipation 
for the case of LD engines operating at MEP. From Eqs. (\ref{compare2}) and 
(\ref{eff1}), we have
\begin{equation}
\frac{x_c}{x_h} = \sqrt{\frac{\Sigma_h(2-\eta^*)}{2\Sigma_c(1-\eta_{\rm C})}},
\end{equation}
which can be solved to give
\begin{equation}
\mathcal{D}_c^{(P_\eta)} = \mathcal{D}_h^{(P_\eta )}  
\left(1-\frac{\eta^*}{2}\right).\label{diss3}
\end{equation}
Comparing Eqs. (\ref{diss1}), (\ref{diss2}) and (\ref{diss3}), it is clear that 
ratio 
of cold to hot dissipation is smallest in the case of Omega function:
\begin{equation}
\frac{\mathcal{D}_c^{(\Omega)}}{\mathcal{D}_h^{(\Omega)}}
< \frac{\mathcal{D}_c^{(P_\eta)}}{\mathcal{D}_h^{(P_\eta)}}<
\frac{\mathcal{D}_c^{(P)}}{\mathcal{D}_h^{(P)}}=1.
\end{equation}
Here, we emphasize that as the ratio of the rates of dissipation at the cold and 
the 
hot ends decreases, the efficiency of
the engine increases, which is clear from the fact that in strong coupling 
limit, engines 
operating at MOF are the most efficient ones
and the engines working in the MP regime are the least efficient 
\cite{Arias2009}.
We also note that in the cases of MP and MOF, the ratio of rates of dissipation 
is independent of dissipation constants $\Sigma_c$ and $\Sigma_h$, whereas for 
MEP it depends 
upon $\gamma$ as the general form of EMEP is a function of $\gamma$. 
\begin{figure}[ht]
 \begin{center}
\includegraphics[width=8.6cm]{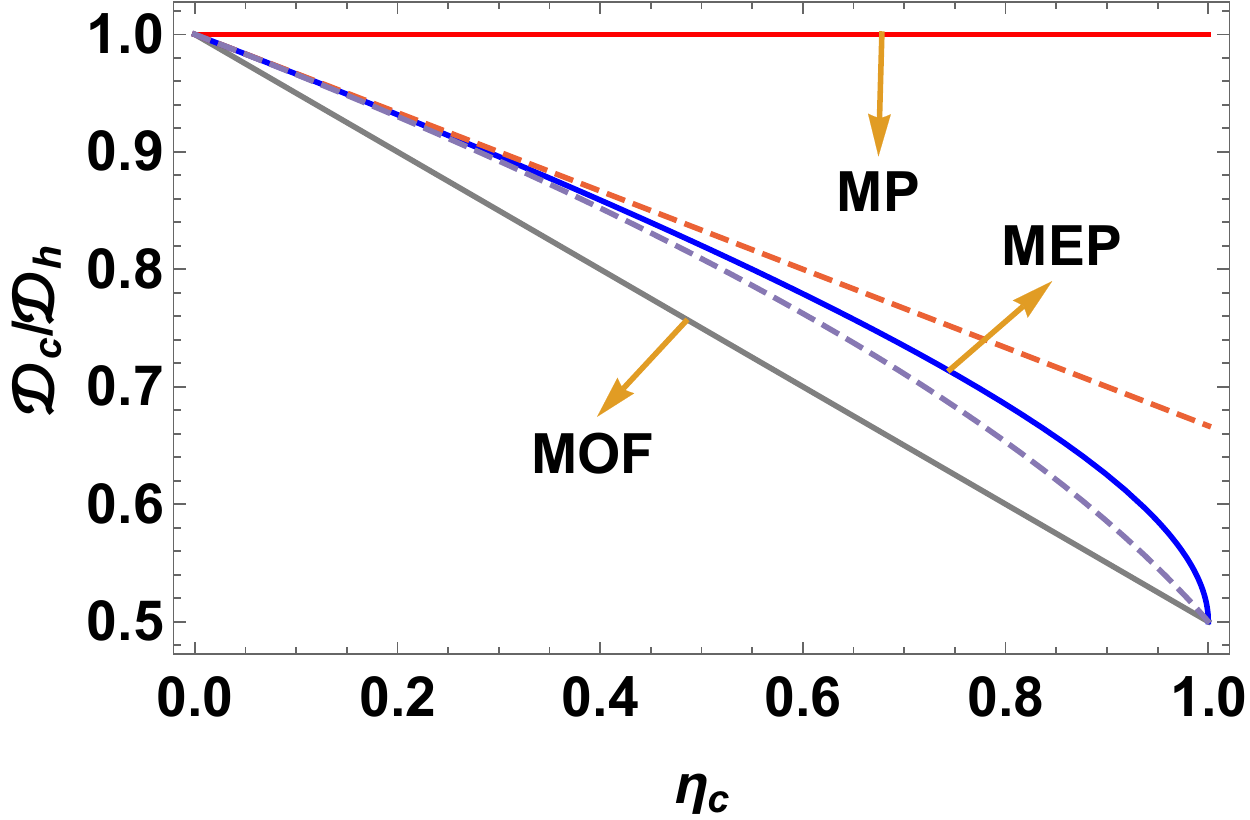}
 \end{center}
\caption{(Color online) Solid lines represent the ratio of the rates of 
dissipation at cold and hot contacts of
the indicated function under symmetric dissipation, $\gamma=1$. 
Dashed upper and lower curves represent the ratio 
$\mathcal{D}_c^{(P_\eta)}/\mathcal{D}_h^{(P_\eta)}$
for the extreme asymmetric dissipation $\gamma\rightarrow \infty$ and 
$\gamma\rightarrow 0$, respectively.}
\end{figure}
\section{Global linear-irreversible principle}
We noted in the above that the bounds on EMEP have also been obtained 
with other models such as the endoreversible model. The similarities and
differences between endoreversible and LD models have been discussed 
recently \cite{Johal2017,Ayala2016,Ayala2017,Tang2018}. While different such 
models assume a particular
functional form or a mechanism for irreversible entropy generation, 
we discuss in the following a 
different formulation that has been recently proposed by one
of the authors \cite{Johal2018} and show that the same lower 
and upper bounds as obtained in Eq. (\ref{bounds}) and (\ref{etasym})
can be obtained using a different optimization scheme. 
In this so-called global linear-irreversible principle (GLIP) framework,
we do not assume stepwise details of the cycle. Rather, 
the validity of LIT is assumed globally, \textit{i.e.}, 
for the complete cycle. Here, the thermal machine  
is considered as an irreversible channel 
with an effective heat conductivity $\lambda$, with an associated passage of
a mean heat $\bar{Q}$ from hot to cold reservoir in the total cycle time 
$\tau$. 
Thereby, the relation between total cycle time $\tau$  and
$\bar{Q}$ is given by \cite{Johal2018}
\begin{equation}
\tau = \frac{\bar{Q}^2}{\lambda\Delta S_{\rm tot}}, \label{cycletime}
\end{equation}
where $\Delta S_{\rm tot} = Q_c/T_c-Q_h/T_h$, is the total entropy generated per 
cycle.
Using the basic definitions and Eq. (\ref{cycletime}), the 
average efficient power is given by
\begin{equation}
P_{\eta} = \eta \frac{W}{\tau} = \frac{\lambda (Q_h-Q_c)^2 }{Q_h\bar{Q}^2}\Delta 
S_{\rm tot} 
=
\frac{\lambda (Q_h-Q_c)^2 
}{Q_h\bar{Q}^2}\left(\frac{Q_c}{T_c}-\frac{Q_h}{T_h}\right). \label{EP2}
\end{equation}
Defining $\nu=Q_c/Q_h$, we can rewrite Eq. (\ref{EP2})
in terms of $\eta_{\rm C}$ and $\nu$:
\begin{equation}
P_{\eta} = \frac{\lambda}{T_c} 
(1-\nu)(\nu+\eta_{\rm C}-1)\frac{{Q}^2_h}{\bar{Q}^2}.\label{EP3}
\end{equation}
Now, in order to optimize the above objective function,
we have to specify the form of $\bar{Q}$ which is assumed
to be a mean value lying in the range $Q_c\leq \bar{Q}\leq Q_h$. We will discuss 
here only 
the extreme cases. Substituting $\bar{Q}=Q_h$ in Eq. (\ref{EP3}) and optimizing 
with respect to $\nu$,
EMEP comes out to be $\eta_-=2\eta_{\rm C}/3$. Similarly, for $\bar{Q}=Q_c$, the 
form of EMEP 
is $\eta_+=(3-\sqrt{9-8\eta_{\rm C}})/2$. Alternately, if we use the geometric 
mean
$\bar{Q}=\sqrt{Q_c Q_h}$, the optimization of Eq. (\ref{EP3}) yields the EMEP as 
in Eq. (\ref{etasym}). 
\section{Conclusions}
We have discussed the efficiency of a LD heat engine 
operating at MEP. In the limit of extremely asymmetric dissipation, we are 
able to obtain the lower and upper bounds on the efficiency of the engine, as 
well
as the expression $\eta_{sym}$ for the symmetric case.
The universal features of EMEP are highlighted. We also note that ratio of 
average 
dissipation rates at cold and hot contacts depends upon $\gamma$, see Eq. 
(\ref{diss3}), 
whereas in the case of MP and MOF, the same ratio is 
independent of $\gamma$, see Eqs. (\ref{diss1}) and (\ref{diss2}). The 
derivation of forms of EMEP,
similar to those obtained for LD Carnot-like engines, using the 
global principle of LIT, confirms the validity of our analysis.

Although the real power plants do not operate in a Carnot cycle, and the 
assumption of low dissipation
may not be valid for them, it is compelling to compare the upper and lower 
bounds with 
the observed efficiencies.
In Fig. 1, we have compared the observed data with the bounds obtained for LD 
engines operating 
at MP and MEP. Although not shown in Fig. 1, it is important to know that the 
area between the lower and 
upper bounds of MOF does not contain any observed data points 
\cite{Holubec2015}, 
whereas five and eight data points respectively lie within the areas bounded by 
the lower and upper bounds 
of EMEP and EMP. However, it is interesting to observe that
the density of points (number of data points per unit area
between the upper and lower bounds for the respective objective
function shown in Fig. 1),
is higher in the case of MEP criterion than for MP. 

\appendix\section{}
Substituting the values of $Q_h$ and $Q_c$ from Eqs. (\ref{heat1}) 
and (\ref{heat2}) into the Eqs. (\ref{opti1}) and (\ref{opti2}) and then adding, 
we have
\begin{eqnarray}
&&T_h  (\Delta S-x_h\Sigma_h) - 2T_c(\Delta S+x_c\Sigma_c)- 2T_c x_c\Sigma_c 
\left[1 - \frac{T_c(\Delta S+x_c\Sigma_c)}{T_h(\Delta S-x_h\Sigma_h)}\right] 
\nonumber
\\
&& +
T_h\left[ 1-\frac{T_c^2(\Delta S+x_c\Sigma_c)^2}{T_h^2(\Delta S-x_h\Sigma_h)} 
\right]
-
x_h\Sigma_h T_h \left[1 -  \frac{T_c^2(\Delta S+x_c\Sigma_c)^2}{T_h^2(\Delta 
S-x_h\Sigma_h)^2}\right] = 0.
\end{eqnarray}
Further writing the above equation in terms of $\eta_{\rm C}=1-T_c/T_h$, we have
\begin{eqnarray}
\Delta S &&-  2 x_h\Sigma_h - 4(1-\eta_{\rm C})x_c\Sigma_c - 2(1-\eta_{\rm 
C})\Delta S    
\nonumber
\\
&& + 
2 (1-\eta_{\rm C})^2 x_c\Sigma_c \frac{\Delta S+x_c\Sigma_c}{\Delta 
S-x_h\Sigma_h} 
+
 \Delta S(1-\eta_{\rm C})^2 \frac{(\Delta S+x_c\Sigma_c)^2}{(\Delta 
S-x_h\Sigma_h)^2}= 0.\label{compare1}
\end{eqnarray}
Solving Eq. (\ref{compare1}) for $x_c$, we have
\begin{equation}
x_c = \frac{1}{\Sigma_c(1-\eta_{\rm C})}\left[\frac{\Delta S^2\eta_{\rm 
C}}{3\Delta S-2x_h\Sigma_h}-x_h\Sigma_h\right].\label{xc1}
\end{equation}
Dividing Eqs. (\ref{opti1}) and (\ref{opti2}) and writing in terms of $\eta_{\rm 
C}$, we get
\begin{equation}
\frac{x_c^2}{x_h^2} = \frac{\Sigma_h}{2\Sigma_c}\left[ \frac{\Delta 
S+x_c\Sigma_c}{\Delta S-x_h\Sigma_h} + \frac{1}{1-\eta_{\rm 
C}}\right].\label{compare2}
\end{equation}
Again solving Eq. (\ref{compare2}) for $x_c$ and writing in terms of $\gamma$, 
we have
\begin{eqnarray}
x_c &=& \frac{1}{4(1-\eta_{\rm C})\Sigma_c(\Delta S-x_h\Sigma_h)}\bigg[\gamma 
x_h^2\Sigma_h^2(1-\eta_{\rm C})  
-
x_h\Sigma_h\sqrt{\gamma(1-\eta_{\rm C})} \nonumber
\\
&& \times \sqrt{
8\Delta S^2(2-\eta_{\rm C}) -8\Delta S x_h\Sigma_h(3-\eta_{\rm C}) 
+
 \gamma x_h^2\Sigma_h^2(1-\eta_{\rm C}) + 8x_h^2\Sigma_h^2} \bigg]             
\label{xc2}
\end{eqnarray}
Eliminating $x_c$ from Eqs. (\ref{xc1}) and (\ref{xc2}), we have the final 
expression for $x_h$ as given by Eq. (\ref{xh1}).
\section{}
Efficiency of the engine is given by:
\begin{equation}
\eta = \frac{W}{Q_h} = 1 - \frac{Q_c}{ Q_h}.
\end{equation}
Using Eq. (\ref{heat1}) and (\ref{heat2}) and writing in terms of $\eta_{\rm 
C}$,
the expression for efficiency becomes
\begin{equation}
\eta = 1 - (1-\eta_{\rm C}) \frac{\Delta S+x_c\Sigma_c}{\Delta S-x_h\Sigma_h} 
\label{eff1}.
\end{equation}
Rearranging the terms in Eq. (\ref{xc1}), we obtain under conditions of MEP
\begin{equation}
(\Delta S+x_c\Sigma_c)(1-\eta_{\rm C}) = \frac{\Delta S^2\eta_{\rm C}}{3\Delta 
S-2x_h\Sigma_h} + \Delta S-x_h\Sigma_h -\Delta S\eta_{\rm C}.\label{eff3}
\end{equation}
Substituting Eq. (\ref{eff3}) in Eq. (\ref{eff1}), we obtain following form of 
efficiency
\begin{equation}
\eta^* = \frac{2\eta_{\rm C}}{3 -2x_h\Sigma_h/\Delta S}.
\end{equation}.
%

\bibliographystyle{apsrev4-1}

\end{document}